\documentclass[superscriptaddress, floatfix, unsortedaddress, aps, pre,
amssymb, amsmath, twocolumn]{revtex4}
\usepackage{amsmath}
\usepackage{graphicx}

\begin{document}

\bibliographystyle{epj}
\title{Anisotropic Elastic Model for Short DNA Loops}

\author{G. Bizhani}
\address{Department of Physics, Sharif University of Technology,
P.O. Box 11365-9161, Tehran, Iran.}

 \author{ N. Hamedani Radja}
\address{Department of Physics, Sharif University of Technology,
P.O. Box 11365-9161, Tehran, Iran.}
\address{Institute for Studies in
Theoretical Physics and Mathematics, P.O.Box 19395-5531, Tehran,
Iran.}

\author{F. Mohammad-Rafiee}
\address{Institute for Advanced Studies in Basic Sciences, P. O.
Box 45195-1159, Zanjan 45195, Iran.}
\address{Physico-Chimie
Th\'{e}orique, UMR 7083, ESPCI 10 rue Vauquelin, F-75231 Paris
Cedex 05, France.}

\author{M. R. Ejtehadi}
\address{Department of Physics, Sharif University of Technology,
P.O. Box 11365-9161, Tehran, Iran.}

\date{\today}

\begin{abstract} Effect of bending anisotropy on a planar DNA loop,
using energy minimization and neglecting entropic effects,  is
studied. We show that the anisotropy  results in
polygonal shape of the loop and increasing the anisotropy makes
the edges sharper. Calculating the energy of such a loop lets us
to find effective persistence length as the geometrical mean of
hard and soft rigidities, which is quite different from harmonic mean
for an unconstrained long DNA.
\end{abstract}

\maketitle
\section{Introduction}

A long time has been passed since a straight double helical
structure had been proposed by Watson and Crick as relaxed
configuration of a DNA, but today, we know it 
can almost never be found in this form in  nature. The macromolecule
is able to play its important and essential role in the life, when
it is interacting with different proteins which force it to bend,
twist, melt and/or pack~\cite{matthews92,schleif92}.

To understand the macromolecule's functionality, DNA has been
examined in different length scales. Single molecule force-extension
studies on $10-100\mu$~m DNAs~\cite{Smith92}, cyclization
experiments on $100-1000$~nm DNAs~\cite{han97} and recent
experiments on sharply bent DNAs in loops with  the lengths less
than  $50$~nm, persistence length of
DNA~\cite{widom04,widom05,saiz05,Vologodskii05} . All these
experiments are of great biological interest. Depending on relevant
length scale of the problem, experimental results are usually
analyzed in two types of theoretical models: variants of the
continuum elastic worm-like chain (WLC) and base-pair steps.

The worm-like chain models are based on the physics of elastic rod
where the energy has a quadratic form in deformations. There are two
approaches to this model. The first one is based on Kirchhoff-like
equations of balance of forces and torques for every segment of the
rod~\cite{mahadevan96,westcott97} and the other one is based on
energy minimization~\cite{moroz98,zhang03}. In these models
statistical mechanics of the rod at non-zero temperature is also
encountered~\cite{yamakawa78}. A force constant or
equivalently a deformation module is defined for bending, twisting,
stretching and their coupling terms.
In the classical WLC model, all cross terms as well as stretching energy
are neglected and the rod is isotropic in bending in different
directions~\cite{shimada84}. So the model is identified with just
two persistence lengths assigned to each  of  bending or twisting
degrees of freedom. Modifications to this model have been made
by encountering twist-stretch coupling, twist-bend coupling and
stretch-bend
coupling~\cite{moroz98,marko94,kamien97,O'Hern98,FarshidPRL05} and
estimating  the free parameters by fitting the theory to
force-extension experimental data. WLC model
has also been used to evaluate loop formation
 probability, $J$-factor~\cite{zhang03,shimada84}.

Looking at microscopic structure of DNA macromolecule suggests that
bending toward the groove is easier than bending toward the
backbone. Though, there is no direct experiment to measure the
anisotropy, there are some theoretical postulates for double
stranded DNA~\cite{sclellman74,zhurkin79,matsumoto02}.
Monte Carlo simulations have also confirmed that B-DNA bends more
easily in the groove direction (roll) than in the backbone direction
(tilt)~\cite{zhurkin91}. It has been encountered in some models for
describing force-extension and DNA cyclization experiments and found
to have no significant improvement to the
results~\cite{moroz98,zhang03,levene86}, though existence
of bending anisotropy affects the existence of twist-stretch
coupling~\cite{O'Hern98}. Balaeff
\textit{et al.} have claimed that bending anisotropy can reduce the
energy of loops in length scale of $76$~bp  by a factor of one
third~\cite{balaeff04}. In a recent experiment an asymmetry in the
periodic behavior of free energy of loop formation as a function of
loop length for $60-100$~bp loops has been detected~\cite{saiz05}.
Fourier analysis of free energy shows two main frequencies. One with
a period of $\sim10.5$~bp due to helical shape of the double strand
and one with a period of $\sim5.6$~bp that might be a result of
bending anisotropy.

Sequence dependence and anisotropy of bending persistence length has
been widely noticed in base-pair steps approaches, in which relative
rotation and displacement of every two segments are defined through
six parameters slide, shift, rise, tilt, roll and
twist~\cite{olson93}. These sequence dependent parameters
for individual base-pair steps have been determined from their
standard deviation in crystal complexes~\cite{olson98}. Theoretical
work has also been done on extracting these parameters from atomic
level parameters via the analysis of Molecular Dynamics (MD)
trajectories~\cite{Gonzales00}. Different simulation studies give
estimations for roll and tilt values. Munteanu \textit{et al.} have
shown that for $3-11$~bp DNAs bending rigidity oscillates with
bending direction and the values of roll is $\sim8-10$ times greater
than the values of tilt~\cite{munteanu98}. Olson \textit{et al.}
estimate the ratio of the hard bending rigidity, $A_1$, to the soft
bending rigidity $A_{2}$ to be between $4$
 and $16$~\cite{olson93} and between $1$ and $5$ in a more
recent study~\cite{olson04}. MD simulations of $17$~bp dsDNAs
gives $A_{1}$ almost twice as $A_{2}$~\cite{Lankas00}. In a stack
of plates simulations, Mergell \textit{et al.} state that spacial
constraints make roll twice as favorable as tilt
\cite{mergell03}. Effect of sequence dependency on persistence
length and also intrinsic curvature on loop formation problem have
been studied analytically by Popov \textit{et al.} which result
in a wide distribution of cyclization probabilities~\cite{popov}.

An effective persistence length, $A$, can be assigned to a rod with
hard and soft  
anisotropic bending rigidities $A_{1}$ and $A_{2}$ (corresponding to
tilt and roll). In an analytical
stack of plates study, done by Olson {\it et al.}, a complicated
dependence of $A$ on detailed couplings, anisotropic constants and
sequence alphabet has been shown for a free DNA~\cite{olson04}.

For a free, long and highly twisted DNA, the effective persistence
length exactly  equals the harmonic mean of soft and hard
rigidities,
\begin{equation}
\label{Eq:harmonic} A=2(\frac{1}{A_{1}}+\frac{1}{A_{2}})^{-1}.
\end{equation}
This can be easily deduced from the equipartition principal. 
The equipartition principle states that the total energy of a  rod is
equal to $\frac{1}{2}k_BT$ times number of its degrees of freedom. 
Bending of a segment of a rod around a principal axis with length
equal to its persistence length can be considered as one degree of
freedom. 
Therefore number of degrees of freedom in hard and soft directions is
simply counted as $\frac{L}{A_1}$ and $\frac{L}{A_2}$ 
and the total energy is found to be $\frac{1}{2}k_BT (\frac{L}{A_1} +
\frac{L}{A_2})$. 
To find the efficient persistence length, the anisotropic rod is
considered as an isotropic one. 
Since the isotropic rod is able to be bent around two similar
directions, its energy is found to be $2(\frac{1}{2}k_BT \frac{L}{A})$ in
which $A$ is the efficient persistence length. 
Having the two estimations equal to each other results
equation~\ref{Eq:harmonic}. This is more
accurately derived by Kehrbaum using an averaging
theory for a non-isotropic elastic rod with high intrinsic
twist~\cite{kehrbaumthesis}. Adding  geometrical constraints on a free
DNA might
affect the above relation.

Here we are going to give an expression for effective persistence
length of  in-plane DNA loops, by energy
minimization, when the entropic effects are neglected. Although an isotropic
WLC loop with length smaller enough than its persistence length is
planar, it is not obvious for the case of anisotropic DNA. In fact
experiments~\cite{amzallag06} and MD simulations~\cite{Lankas06}
show that DNA minicircles are not completely, but are almost planar.

The rest of paper is organized as follows: Section 2 describes the
model that is used in studying the short DNA loops, followed by the
presentation of the results in Section 3. Finally, Section 4
concludes the paper, while the overestimation of deviation of DNA
local twist from its mean value appears in Appendix.

\section{The Model}
The anisotropic elastic model of DNA represents the macromolecule as
an elastic rod of length $L$, parameterized by arclength $s$. As the
double-strand is ribbon-like, it is \ anisotropic in bending around
two different directions. On the other hand, DNA is twisted, so
while bending around a fixed axis, it should bend around each of the
two directions, periodically. It is energetically preferable for the
DNA to bend more around the ``soft'' axis and less around the
``hard'' one. To reduce elastic energy, the DNA also may modify its
twist to have more soft bending along its planar path.

For a planar DNA loop, the tangent unit vector $\hat{t}$ and the
twist angle $\psi$  at each point contain enough information to
parameterize DNA conformation. Actually an anisotropic bent DNA has
a higher tendency to go out of the bending plane which depends on
the value of anisotropy $(A_{1} - A_{2})/(A_{1} + A_{2})$. For short
loops, this 
tendency decreases as the  anisotropy becomes smaller~\cite{farshid03}.

Elastic energy of anisotropic rod is
\begin{equation}
\label{eq:E_A} E=\frac{1}{2}k_{B}T \int_{0}^{L}\left[
A(s)|\dot{\hat{t}}\,|^2+C(\dot{\psi} - \omega_0)^{2} \right]ds,
\end{equation}
where $A(s)$ is the local bending rigidity, $C$ is the twist
rigidity, $\omega_{0}$ is the spontaneous twist of the helix, and
the dots indicate derivatives with respect to $s$. The first term is
the usual elastic term, with $s$ dependent persistence length $A(s)$
and the second term is the energy needed to over-(under-)twist the
DNA, which may be implied by the boundary conditions on $\psi$ or
the anisotropic effects. It should be noted that in this model $s$
dependence of $A$ is not because of sequence dependence, but it is
due to rotation of soft and hard axes. As we will show later, this
is a source of twist-bend coupling, although there is no explicit
coupling term in the Hamiltonian.

DNA bending could be decomposed into two principal axes $\hat{e}_1$
and $\hat{e}_2$, attached to the DNA.  The hard one, $\hat{e}_1$ is
perpendicular to double strand's local plane, and  the soft  one,
$\hat{e}_2$, is defined to lie in the local plane of the double
strand, and to be perpendicular to both strands. Due to the helical
structure of DNA, $\hat{e}_1$ and $\hat{e}_2$ rotate with the helix.
Since $\dot{\hat{t}}$ is perpendicular to $\hat{t}$ axis and lies
in the $\hat{e}_1-\hat{e}_2$ plane, by its decomposition in our
coordinate system we obtain
\begin{eqnarray}
\label{eq:E_A1A2} E &=& \frac{1}{2} k_{B}T  \int_{0}^{L} \left[(A_1
\sin^{2}\psi +A_2 \cos^{2}\psi ) \, \dot{\theta}^{2} \right.
\nonumber
\\
 && ~~~~~~~~~~~~~ \left. + ~ C(\dot{\psi} - \omega_0)^{2}\right]ds ,
\end{eqnarray}
where $A_1$ and $A_2$ are constant bending
rigidities about the rotating axes, $\hat{e}_1$ and $\hat{e}_2$, and
$\theta$ is the angle between $\hat{t}$ and a fixed, arbitrary
direction in loop's plane. Using Euler-Lagrange equation and applying
corresponding boundary conditions, we are able to find $\theta(s)$ and
$\psi(s)$ as well as DNA's shape and bending energy of the loop. The
integral form of closed loop or ``ends meeting'' condition is
\begin{equation}
\label{eq:EndMeet} \int_{0}^{L}\sin\theta(s)\,\mathrm{d}s =
\int_{0}^{L}\cos\theta(s)\, \mathrm{d}s = 0.
\end{equation}
Also because the DNA strands are antiparallel they can not switch in
the ends and they should bind in phase, then we have
$\psi(L)-\psi(0)=2k\pi$ with $k$ being an integer that is the number
of full turns of the helix. On the other hand to avoid any
singularity in tangent vector changes we consider a constraint on
$\theta$ values in the ends by $\theta(L)-\theta(0)=2\pi$ (no self
crossing).

\section{Results}

Integrating Euler-Lagrange Equations of (\ref{eq:E_A1A2})
 results following
equations of motions:
\begin{equation}
\label{eq:EL1} \dot{\theta}=\frac{\gamma}{(A_1+A_2) - (A_1-A_2)\cos 2\psi},
\end{equation}

\begin{equation}
\label{eq:EL3} \dot{\psi}^{2}=-\frac{2\gamma^2}{C} \,
\frac{1}{(A_1+A_2) - (A_1-A_2)\cos 2\psi}+\beta,
\end{equation}
where integral constants of $\gamma$ and $\beta$ should be
determined from the boundary conditions. As it is seen in above
equations, $\psi$ depends on $s$, as on the DNA's local
bending in a
complicated form. This twist-bend coupling is a direct result of
anisotropy of the model and vanishes in the case of isotropic rods.
 Even if  $A_1$ and $A_2$ differ by one order of magnitude,
which is the case of our double stranded DNA, the coupling is very
weak and $s$ dependence of $\dot{\psi}$ is negligible. Indeed,
numerical studies show that the relative variations of $\dot{\psi}$
is less than one percent,  even if two bending rigidities differ by
two orders of magnitude~\cite{farshid03}. This leads us to consider
homogeneous twist along the DNA and set $\psi(s)= \omega s$, in
which $\omega=2 k \pi /L$ is found by applying the ``ends in phase''
boundary condition.
This is more discussed in Appendix. The above approximation
decouples torsional part of energy from its bending part.


Integrating~(\ref{eq:EL1}) with considering above approximation gives
\begin{eqnarray}
\theta(s)=\theta(0)&+&\frac{\gamma}{\omega\sqrt{A_1
A_2}} \\
&\times&\left( \tan^{-1}\left(\sqrt{\frac{A_1}{A_2}}\tan(\omega
s)\right) +\pi\left[\frac{1}{2}+\frac{\omega
s}{\pi}\right]\right),\nonumber
\end{eqnarray}
where the bracket means ``integer part''. This term is added to
get rid of discontinuity in the $\tan^{-1}$ function. Without lack
of generality, we set $\theta(0)=0$ and therefore the condition on
total bending results $\theta(L)=2\pi$. Applying this condition simply
lets us to fix $ \gamma = (2\pi/L) \sqrt{A_1
A_2} $. These would yield the functional form of $\theta(s)$ and
loop's shape as
\begin{equation}
\theta(s)=\frac{1}{k}\left(\tan^{-1}
\left(\sqrt{\frac{A_1}{A_2}}\tan(\omega s) \right) +\pi
\left[\frac{1}{2}+\frac{\omega s}{\pi} \right]\right).
\end{equation}
The above solution automatically satisfies the ``ends meeting''
condition~(\ref{eq:EndMeet}). $k$ in above equation counts the
number of turns of helix along the loop. To have minimum torsional
energy it should be fixed to the closest integer value to
$\omega_{0}L/2\pi$, so $k=round(\omega_{0}L/2\pi)$. Thus the DNA is
undertwisted in case
$\Delta\psi = \omega_{0} L - 2 k \pi < 0$
and is overtwisted otherwise.

\begin{figure}
\includegraphics[width=1\columnwidth]{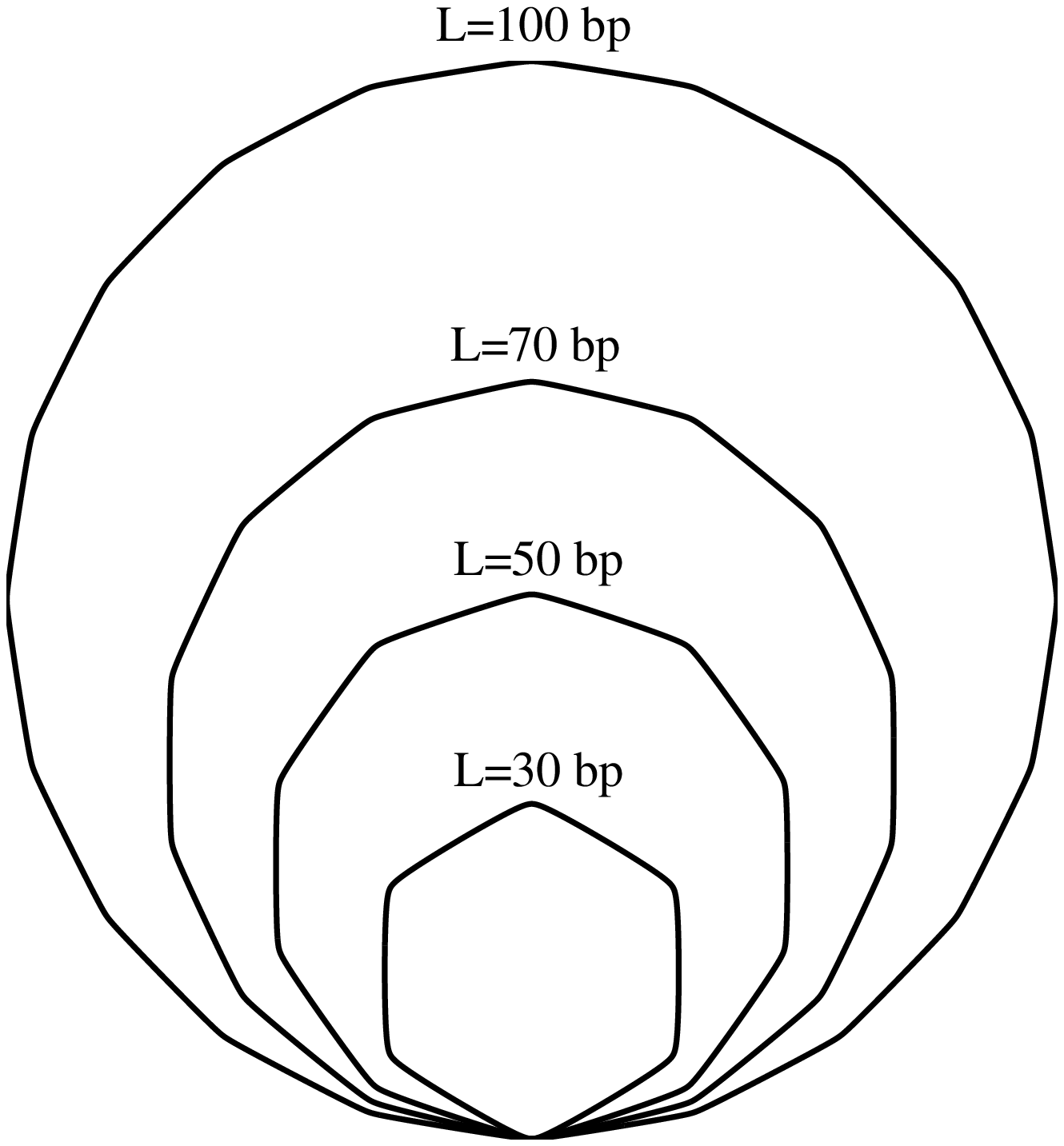}
\includegraphics[width=1\columnwidth]{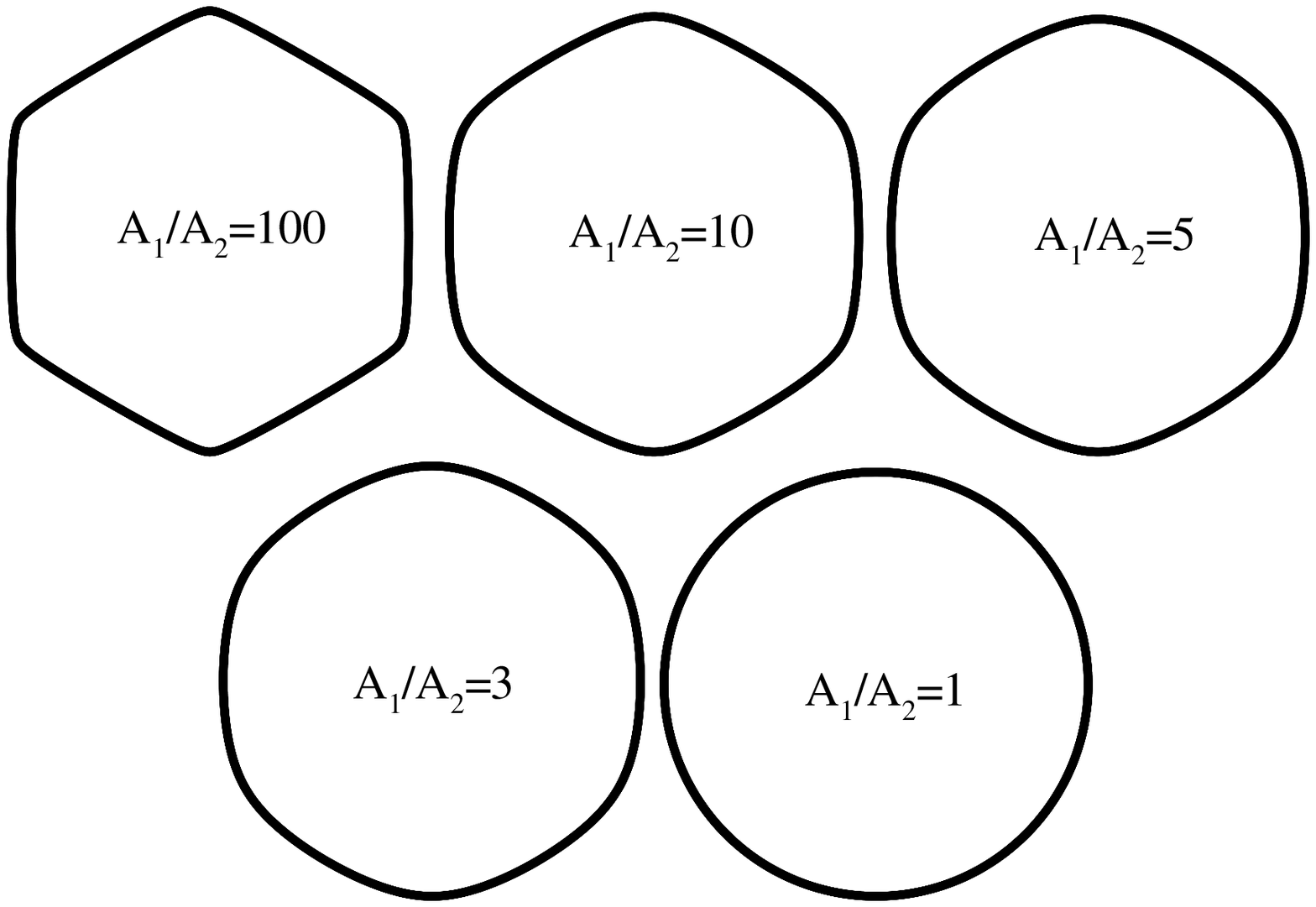}
\caption{\textbf{Above}, Loop shapes for different DNA lengths. An
exaggerated ratio of $\frac{A_{1}}{A_{2}}=100$  for all illustrations has been
used. \textbf{Below}, Effect of
anisotropy on sharpness of polygonal edges. A loop length of
 $L=9.6$~nm for illustrations has been used. In all figures $\omega=1.8$~
nm$^{-1}$ (see text).} \label{fig:conformation} 
\end{figure}

In the case $A_1 >> A_2$ for a closed DNA loop, bending is not
homogeneous and it is localized in ``soft" parts. Here the loop
looks more like a polygon rather than a circle. This is the direct
effect of bending anisotropy. The
shape of looped DNA for anisotropic model is given in
figure~\ref{fig:conformation}. Increasing the ratio of $A_1/A_2$,
the polygonal shape is more visible (e.g. for $A_1/A_2>50$, the DNA
loop will have sharp edges at the soft points).  In a full helix turn
the DNA meets the soft axis of 
rotation two times when it bends in plane, thus the number of
polygon edges equals $2k$. As the number of edges ($\sim$ helical
turns) increases by length of the loop, the polygonal shape of the
loop is less visible for larger DNAs.

To find the energy of the loop, we read $(A_1\sin^{2}\psi +
A_2\cos^{2}\psi)\, \dot{\theta}=\gamma$ from~(\ref{eq:EL1}) and
substitute it in~(\ref{eq:E_A1A2}),
\begin{equation}
\label{eq:E_gamma} E=\frac{1}{2}k_{B}T \int_{0}^{L} \left(\gamma
\dot{\theta} + C(\dot{\psi} - \omega_0)^{2} \right) ds.
\end{equation}
As $\int_{0}^{L} \dot\theta ds = 2\pi$ and $\dot\psi$ and $\gamma$
are constant, the elastic energy is
\begin{equation}
E=\frac{2\pi^{2}}{L}k_{B}T \sqrt{A_1 A_2}  + \frac{k_{\rm B}T}{2L} C
\Delta\psi^2.
\end{equation}

The second term is the twist energy which is due to the
over-(under-)twist implied by the ``ends in  phase''  condition on
DNA loop. Because $k$ is a step function of L, this term leads to a
well known oscillatory behavior which is damping by an $L^{-1}$
factor.
In analogy with the bending energy of a circular loop,
$(2\pi^{2}k_{B}T A/L)$, we can read the effective persistence length
of in-plane small loops as $\sqrt{A_1 A_2}$. This is different from
the effective persistence length of anisotropic DNA in larger
scales (equation~(\ref{Eq:harmonic})).


\section{Conclusion}

We studied the effect of bending anisotropy on planar DNA loops
using energy minimization and neglecting entropic effects under
constraints of parallel and in phase ends. Bending anisotropy causes
anisotropy in curvature and twist. However, anisotropy induced a
twist-bend coupling to the model even in the lack of explicit
appearance of such coupling in the Hamiltonian, though, it is small
enough to be neglected in the calculations.  The bending anisotropy
results in polygonal shape of the loop. Increasing the anisotropy
makes the edges sharper where the number of helical turns and hence
the number of edges of the polygon grows with loop length.

Energy of such a loop includes an oscillating term for twist energy
and a bending term similar to that of anisotropic loop with an
effective persistence length $A=\sqrt{A_{1}A_{2}}$, which is
different from harmonic mean of $A_{1}$ and $A_{2}$
(equation~\ref{Eq:harmonic}) for an unconstrained long DNA. As the
geometric mean of $A_1$ and $A_2$ for $A_1 \neq A_2$ is always larger
than their harmonic mean, the results shows that considering
anisotropy in DNA bending rigidities is not able to  explain large
loop formation probability of DNA minicircles. Thus it seems other 
efforts, as like as bubble formation~\cite{YanMarkoPRL04}, effect
of sequence dependence~\cite{popov}, or generalized semiflexible
model~\cite{Wiggins2006} are more successful in this direction.

\appendix
\section*{Appendix}
We are going to give a simple numeric estimation for deviation of
$\dot\psi$ from its mean value. Since $\cos 2 \psi$ varies between
$-1$ and $1$
the maximum value of the $\psi$ dependent term  (first term) of
equation~(\ref{eq:EL3}) is
\begin {equation}
\max(x) =\frac{-\gamma^2}{C}
\frac{1}{A_{2}}=\frac{-4\pi^{2}A_{1}A_{2}}{C A_{2}L^{2}}.
\end{equation}
Applying the typical values for a DNA minicircle (i.e.
$\omega=1.8$~nm$^{-1}$, $C=75$~nm, $L=94$~bp and $bp=0.34$~nm)
following by rough estimations, $A_{1}=100$~nm and $A_{2}=33$~nm
(i.e $A_1/A_2=3$ with a harmonic mean of 50~nm), result a $\max(x)
\approx 0.016\,\omega^{2}$. So the deviation of $\dot\psi^2$ is
small comparing  to the mean value and therefore $\dot\psi^{2}\approx
\beta \approx \omega^{2}$. Now finding $(\max-\min)$ of both sides
of equation~(\ref{eq:EL3}), we have:

\begin{equation}
2 \dot\psi \Delta\psi=
\frac{\gamma^{2}}{C}(\frac{1}{A_{2}}-\frac{1}{A_{1}})=\frac{4
\pi^{2}}{L^{2}C}(A_{1}-A_{2})
\end{equation}
and as $\dot\psi\approx\omega$
\begin{equation}
\frac{\Delta\dot\psi}{\omega}=\frac{4 \pi^{2}(A_{1}-A_{2})}{L^{2} C
\omega^{2}}\approx1\% .
\end{equation}
It should be noted that above calculation is an overestimation  of the
range of $\dot\psi$ changes. As it was mentioned in the text, numerical
study shows less than one percent fluctuation in the value even with
considering two rigidities different by two orders of magnitude. That
means, considering $\omega$ as a constant ($\psi=\omega s$),
is not a rude approximation.

{\bf Acknowledgement:} We would like to thank R. Golestanian and
B. Eslami-Mossallam for useful discussions and their valuable comments.  

\bibliography{int}
\end{document}